\documentclass[12pt]{article}

\usepackage{amsmath}
\usepackage{amssymb}
\usepackage{verbatim}

\makeatletter

   \@addtoreset{equation}{section}
\makeatother

\begin{document}

\vskip 12mm

\begin{center} 
{\Large \bf  Higher Spins as Rolling Tachyons in Open String Field Theory}
\vskip 10mm
{ \large  Dmitri Polyakov$^{a,b,}$\footnote{email:polyakov@scu.edu.cn;polyakov@sogang.ac.kr} $\footnote{}$ 

\vskip 8mm
$^{a}$ {\it  Center for Theoretical Physics, College of Physical Science and Technology}\\
{\it  Sichuan University, Chengdu 6100064, China}\\
\vskip 2mm

$^{b}$ {\it Institute of Information Transmission Problems (IITP)}\\
{\it  Bolshoi Karetny per. 19/1, Moscow 127994, Russia}\\

}
\end{center}

\vskip 15mm

\begin{abstract}

We find a simple analytic solution in open string field theory which, in the on-shell limit,
generates a tower of higher spin vertex operators in bosonic  string theory. The solution
is related  to irregular off-shell vertex operators for Gaiotto states. 
The wavefunctions for the
irregular vertex operators are described by equations  following from the cubic effective 
action for generalized
rolling tachyons, indicating that the evolution from flat to collective higher-spin background
 in string
field theory occurs according to cosmological pattern.
 We discuss the relation between nonlocalities of the rolling tachyon action and those 
of higher spin interactions.

\end{abstract}

\vskip 12mm

\setcounter{footnote}{0}

\section{\bf Introduction} 

Higher spin fields in AdS space-time are known to be a crucial ingredient of $AdS/CFT$ correspondence,
as most of the composite operators on the conforml field theory (CFT) side are holographically dual to 
the higher spin modes. Perhaps the best known example of such a correspondence is the one between 
the higher derivative gauge invariant operators in the $O(N)$ vector model and 
the symmetric higher spin frame-like fields in $AdS_4$ \cite{pol, kp, ss, mvf, mvs, mvt}.  But 
generically, any operator on the CFT side
carrying multiple space-time tensor indices, is expected to be dual to some higher-spin field with 
mixed symmetry.
On the other hand, the higher spin holography implies that any correlator in boundary CFT  is 
reproduced by the worldsheet
correlators of vertex operators in string theory, i.e. any gauge-invariant observable on the 
CFT side has its dual vertex operator in 
AdS string theory. For simple some simple CFT/gauge theory observables such a correspondence is 
straightforward.
For example, the
 string counterpart of $Tr(F^2)$ in super Yang-Mills theory is the vertex operator $V_\phi$ of 
a dilaton in string theory,
while the stress-energy tensor $T_{mn}$ corresponds to the graviton's vertex operator $V_{mn}$, 
polarized along the $AdS$ boundary.
One problem with checking this conjecture on the string theory side is that we know little 
about $AdS$ string dynamics beyond
semiclassical approximation. 
This is related to the fact that the first-quantized string theory is background-dependent.

In addition, the higher spin interactions (at least beyond the quartic order) are known to be highly 
nonlocal,
while the standard low-energy effective actions, stemming from vanishing $\beta$-function constraints 
in the
first-quantized theories are typically local. This altogether suggests that the second-quantized 
formalism of
the string field theory \cite{witsft}
may be a more adequate formalism to approach the higher spin holography from the string theory side,
especially given
the formal similarity of background-independent string field theory (SFT) equations of motion 
\cite{witsft},
and Vasiliev's equations in the unfolding formalism \cite{mvuf, mvus, mvut, mvuft} .
This naturally poses a question of how the vertex operator to CFT observable correspondence may be 
extended off-shell,
in particular involving the higher spin modes.
This does not seem to be obvious.
For example, consider a composite operator given by the $N$'th power of the stress-energy tensor in CFT:

\begin{equation}
T^n\sim{T_{m_1n_1}}...T_{m_Nn_N}
\end{equation}

In the gravity limit, this operator must be a dual of a certain field of spin $2N$ in $AdS$ with mixed 
symmetries.
But what is the vertex operator description of such an object in string theory?
To answer this question, one has to take the colliding limit
of $N$ graviton vertex operators in string  theory. Taking such a limit does not look simple and
must not be confused with the normal ordering. Instead, in order to reproduce the correlation 
functions correctly in such a
limit, one has to retain $all$ the terms, up to $all$ orders of the operator product expansion (OPE), 
as the operators
are colliding  at the common point.
Such a limit is well-known in the matrix model formulations of Liouville and Toda theories and plays an 
important
role in extending the $AGT$ conjecture to Argyres-Douglas type of supersymmetric gauge theories with 
asymptotic freedom.
The result is given by rank $N-1$ irregular Gaiotto-BMT (Bonelli-Maruyoshi-Tanzini) states 
\cite{gaiof, agts, agtt, chaihos, chaihot, chaiho, gaios}
These states extend the context of the primary operators in CFT and lead to special representations of 
Virasoro algebra, being the simultaneous eigenstates of $N+1$ Virasoro generators:

\begin{eqnarray}
L_n |U_N>=\rho_n |U_N>
(N\leq{n}\leq{2N})
\nonumber \\
L_n |U_N>=0 (n>2N)
\end{eqnarray}

In the previous work \cite{chaiho} 
(see also \cite{nagoya, lefloch} with the related issues addressed)
 it was shown that the irregular states admit the following
irregular vertex operator representation in terms of Liouville or Toda fields
:

\begin{eqnarray}
|U_N>=U_N|0>
\nonumber \\
U_N=e^{{\vec{\alpha_0}}{\vec{\phi}}+\sum_{k=1}^N{\vec{\alpha}}_k\partial^k{\vec{\phi}}}
\end{eqnarray}
where ${\vec{\phi}}=\phi_1,...\phi_D$ is either $D$-component Toda field or (in the context of the
 present paper)
parametrize the coordinates of $D$-dimensional target space in bosonic string theory.
The  ${\vec{\alpha}}_k$ parameters are related to the Virasoro eigenvalues (1.2) according to \cite{chaiho}

\begin{eqnarray}
\rho_n=-{{1}\over{2}}\sum_{k_1,k_2;k_1+k_2=n}{\vec{\alpha}}_{k_1}{\vec{\alpha}}_{k_2}
\end{eqnarray}
In case of $D{\geq}2$, the irregular states, apart from being eigenvalues of positive Virasoro generators,
are also the eigenstates of positive modes $W_n^{(p)}$ of the $W_n$-algebra currents 
($3\leq{n}\leq{D+1}$) 
where
\begin{eqnarray}
W_n^{(p)}=\oint{{dz}\over{2i\pi}}z^{p+n-1}W_n(z)
\end{eqnarray}
where $W_n$ are the spin $n$ primaries and $(n-1)N\leq{p}\leq{nN}$.
so that
\begin{eqnarray}
W_n^{(p)}U_N=\rho_n^pU_N
\end{eqnarray}
and $\rho_n^p$ are degree $n$ polynomials in the components of ${\vec{\alpha}}$.
Note that, while the maximal possible rank $n$ is always 
at least $D+1$, for higher dimensions ($D>5$)  it is also possible to have  the higher ranks
$n>D+1$  as well. In general case, the upper bound on $n$ is in fact related to a rather complex problem 
in the partition 
theory. Namely, the maximal rank is given by the maximal number $n_{max}$ for which the inequality
\begin{eqnarray}
\sum_{k=1}^{n_{max}}{{(k+D-1)!\kappa(n_{max}|k)}\over{k!}}-\sum_{q=1}^{{n_{max}}-1}\sum_{k=1}^q
{{(k+D-1)!\kappa(n_{max}|k)}\over{k!}}-(D-1)!\geq{0}
\end{eqnarray}
where $\kappa(n|k)$ is the number of ordered partitions of $n$ with the length $k$:
$n=p_1+....+p_k;0<p_1\leq{p_2}...\leq{p_k}$.

The objects (1.3) are obviously not in the BRST cohomology and are off-shell (except for the regular case
${\vec{\alpha}}_k=0;k\neq{0}$ but make a complete sense in background-independent open string field theory.
On the other hand,  the $U_N$-vertices are related to the onshell vertex operators for the 
higher-spin fields.
That is,  $U_N$ is the generating vertex for the higher-spin operators through

\begin{eqnarray}
V_{h.s.}=\sum_{s,\lbrace{k_1},...,k_s\rbrace}{H^{\mu_1...\mu_s}}({\vec{\alpha}}_0)
{{\partial^s(cU_N)}\over{\partial{\alpha_{k_1}^{\mu_1}}...\partial\alpha_{k_s}^{\mu_s}}}
|_{{\vec{\alpha}}_k=0;k\neq{0}}
\end{eqnarray}
where $c$ is the $c$-ghost, $H^{\mu_1...\mu_s}({\vec{\alpha}}_0)$ are the higher spin $s$ fields in the
 target space with masses
$m={\sqrt{2(k_1+...+k_s-1)}}$, at the momentum ${\vec{\alpha}}_0$
with all the due on-shell constraints on $H$ to ensure the BRST-invariance.
Thus the correlation functions (irregular conformal blocks) of the $U_N$-vertices particularly encode the
 information about the higher-spin
interactions in string theory.
At nonzero ${\vec{\alpha}}_k$ the $U_N$ vertices generate the off-shell extensions of higher-spin 
wavefunctions, which can be studied using the string field theory techniques.
A question of particular interest, studied in this work, is to find the higher spin wavefunction 
configurations
in terms of irregular vertex operators, solving
SFT equations of motion analytically. In case if all ${\vec{\alpha}}_k=0$, except for ${\vec{\alpha}}_0$,
the $U_N$ vertex becomes a tachyonic primary. 
Then , multiplied by the space-time tachyon's wavefunction $T({\vec{\alpha}}_0)$
$\Psi=c\int{d^D\alpha_0}T({\vec{\alpha}}_0)U_N$ is an elementary solution of of string field theory 
equations:
$Q\Psi+\Psi\star\Psi=0$ provided that  $T$ satisfies the vanishing tachyon's $\beta$-function constraints
$\beta_T=({1\over2}\alpha_0^2-1)T+const\times{T^2}=0$ in the leading order of string perturbation theory.
This solution is elementary as  it describes the $perturbative$ background change by a tachyon.
In the case of ${\vec{\alpha}}_k{\neq}0$ things become far more interesting.
The wavefunction in the string field
$\Psi=c\int{d^D{\vec{\alpha}}_0}...{d^D{\vec{\alpha}}_N}T({\vec{\alpha}}_0,...,{\vec{\alpha}}_N)U_N$
can now be regarded  as a generating wavefunction for higher spin excitations in string field 
theory
with the SFT solution constraints on $T$ now related to nonperturbative background change due
to higher spin excitations and the effective action on $T$ holding the keys to higher spin 
interactions
at all orders, just like the well-known Schnabl's analytic solution  \cite{schnabl}
describes the 
physics around the minimum of nonperturbative tachyon potential
(that would be calculated up to all orders, from the string perturbation theory point of view)
 The rest of this letter is organized as follows.
In the Section 2 we explore the CFT properties of irregular vertex operators, including the
behavior under finite conformal transformations, relevant to the SFT equations of motion.
In the present work, we particularly concentrate on the rank 1 and search for the SFT 
analytic solutions
in the form:
\begin{eqnarray}
\Psi=c\int{d^D}{\vec{\alpha}}\int{d^D}{\vec{\beta}}T({\vec{\alpha}},{\vec{\beta}})e^{{\vec{\alpha}}
{\vec{\phi}}+
{\vec{\beta}}\partial{\vec{\phi}}}
\end{eqnarray}
We find that, in the leading order, $\Psi$ is an analytic solution if $T$ satisfies the 
constraints, given by
 equations of motion
described by  the nonlocal effective action for generalized rolling tachyons. The nonlocality 
structures are 
controlled by the SFT worldsheet correlators and by the conformal transformations of the 
irregular blocks.
The solution in particular provides a nice example of how the star product in string field theory
 translates into the Moyal product in
the analytic SFT solutions.
Although we explicitly concentrate on rank one case in this letter, the same structure appears to
 persist for higher irregular ranks as well.
In the discussion section, we investigate the physical meaning and significance of the solution, 
found in this work.
In particular, we relate  the nonlocalities in the noncummutative rolling tachyon structures to those of
 interacting higher-spin theories, 
as in the context of our calculation $T({\vec{\alpha}},{\vec{\beta}})$ is the simplest generating function
for collective higher-spin excitations and the nonlocalities in the effective action are naturally 
translated into those for
the interactions of higher spins. 
The effective equations of motion for  $T({\vec{\alpha}},{\vec{\beta}})$ constitute the 
nonperturbative generalization
of $\beta$-function equations in perturbative string theory, and possibly
can be related  to nonlocal field redefinitions for Vasiliev's
invariant functionals \cite{mvs}.
We also comment on connections between the ${\vec{\beta}}$-coordinate of the solution to 
T-duality and the 
doubling of the space-time and comment on possible relations to double field theory formalism.

\begin{center}
\section{\bf Irregular Vertices as String Field Theory Solutions: Rank 1 Case}
\end{center}

We start with the transformation properties of the irregular vertices under the conformal transformations
 $z\rightarrow{f(z)}$, necessary to
compute the correlators in string field theory.
Straightforward application of the stress tensor to the rank one irregular vertex gives infinitezimal 
conformal transformation:

\begin{eqnarray}
\delta_\epsilon{U_1}({\vec{\alpha}},{\vec{\beta}})
=\lbrack\oint{{dw}\over{2i\pi}}\epsilon(w)T(w);{U_1}({\vec{\alpha}},{\vec{\beta}},z)\rbrack
\nonumber \\
=\lbrace{1\over{12}}\partial^3\epsilon\beta^2+{1\over2}\partial^2\epsilon({\vec{\alpha}}{\vec{\beta}})
+\partial\epsilon({1\over2}\alpha^2+{\vec{\beta}}{{\partial}\over{\partial{\vec{\beta}}}})
+\epsilon\partial_z\rbrace
{U_1}({\vec{\alpha}},{\vec{\beta}},z)
\end{eqnarray}
It is not difficult to obtain the finite transformations for $U_1$, by 
integrating the infinitezimal transformations (2.1).

\begin{eqnarray}
{U_1}({\vec{\alpha}},{\vec{\beta}},z)=e^{{\vec{\alpha}}{\vec{\phi}}+{\vec{\beta}}\partial{\vec{\phi}}}
\rightarrow
({{df}\over{dz}})^{{\alpha^2}\over2}
e^{{\vec{\alpha}}{\vec{\phi}}+{{df}\over{dz}}{\vec{\beta}}\partial{\vec{\phi}}
+({\vec{\alpha}}{\vec{\beta}}){{d}\over{dz}}log({{df}\over{dz}})+{1\over{12}}S(f;z)}
\end{eqnarray}
where
$S(f;z)$ is the Schwarzian derivative.

It is straightforward to generalize this result to transformation laws for the irregular vertices of 
arbitrary ranks. 
For the arbitrary rank N the BRST and finite conformal transformations  for the irregular vertices have
 the form:
\begin{eqnarray}
\lbrace{Q, cU_N}\rbrace=
\lbrace\oint{{dz}\over{2i\pi}}(cT-bc\partial{c}); ce^{i\sum_{q=1}^N{\vec{\alpha_q}}\partial^q{\vec{\phi}}}
\rbrace
\nonumber \\
={1\over2}\sum_{q_1=0}^N\sum_{q_2=0}^N{{q_1!q_2!}\over{{(q_1+q_2+1)}!}}
({\vec{\alpha_{q_1}}}{\vec{\alpha_{q_2}}}):\partial^{q_1+q_2+1}ccU_N
\nonumber \\
+i\sum_{q=1}^N\sum_{p=1}^{q-1}{{q!}\over{p!(q-p)!}}\partial^{q-p}cc({\vec{\alpha}}_q
{{\partial\over{\partial{\vec{\alpha}}_{p+1}}}})U_N
\end{eqnarray}
and

\begin{eqnarray}
U_N\rightarrow({{df}\over{dz}})^{{\alpha_0^2}\over2}e^{-\sum_{q_1,q_2=1}^NS_{q_1|q_2}(f;z)
+i\sum_{q=2}^N\sum_{k=1}^{q-1}\sum_{l=1}^k{{(q-1)!}\over{k!(q-1-k)!}}
{{d^{q-k}f}\over{dz^{q-k}}}B_{k|l}(\partial{f}...\partial^{k-l+1}f)
({\vec{\alpha}}_q\partial^{l+1}{\vec{\phi}})}
\nonumber \\
\times
e^{\partial^nf({\vec{\alpha}}_q\partial{\vec{\phi}})}
\end{eqnarray}

where
$S_{q_1|q_2}(f;z)$ are the generalized Schwarzian derivatives of the rank $q_1+q_2$,
given by

\begin{eqnarray}
S_{q_1|q_2}(f;z)={{1}\over{(q_1+q_2)!}}B^{(q_1+q_2)}
({{d\over{dz}}}log{{df}\over{dz}};...{{d^{q_1+q_2}}
\over{dz^{q_1+q_2}}}log{{df}\over{dz}}
\nonumber \\
-{{q_1+q_2+1}\over{(q_1+1)!(q_2+1)!}}
B^{(q_1)}({{d\over{dz}}}log{{df}\over{dz}};...{{d^{q_1}}\over{dz^{q_1}}}log{{df}\over{dz}})
B^{(q_2)}({{d\over{dz}}}log{{df}\over{dz}};...{{d^{q_2}}\over{dz^{q_2}}}log{{df}\over{dz}})
\end{eqnarray}

where 
$B^{(q)}(\partial{f}...\partial^q{f})$ are the Bell polynomials in the derivatives of  ${f(z)}$
defined according to
\begin{eqnarray}
B^{(q)}(\partial{f}...\partial^q{f})=q!\sum_{k=1}^q\sum_{q|p_1...p_k}
{{\partial^{p_1}f...\partial^{p_k}f}\over{p_1!...p_k!\lambda_{p_1}!...\lambda_{p_k}!}}
\end{eqnarray}
where $q=p_1+...p_k;0<p_1\leq{p_2}...\leq{p_k}$ are the length $k$ ordered partitions
of $q$ and $\lambda_p$ is the multiplicity of an element $p$ of the partition.
Given the transformation rules for the irregular vertices, it is now 
straightforward to compute the correlators relevant to
the open string field theory equations of motion.
The evaluation of the kinetic term with the string field of the form (1.9) leads to

\begin{eqnarray}
<<Q\Psi(0)\star\Psi(0)>>=<Q\Psi(0)I\circ\Psi(0)>
\nonumber \\
=lim_{w\rightarrow\infty}
\int{d^D\alpha_1}\int{d^D\beta_1}\int{d^D\alpha_2}\int{d^D\beta_2}T({\vec{\alpha}}_1,{\vec{\beta}}_1)
yT({\vec{\alpha}}_2,{\vec{\beta}}_2)w^{\alpha_2^2-1}e^{-{{{\vec{\alpha_2}}{\vec{\beta_2}}}\over{z}}}
\nonumber \\ 
\times\lbrack
({1\over2}\alpha_1^2-1+{\vec{\beta}}{{\partial}\over{\partial{\vec{\beta}}}})<\partial{c}
ce^{i{\vec{\alpha}}_1{\vec{\phi}}+i{\vec{\beta}}_1\partial{\vec{\phi}}}
(0)ce^{i{\vec{\alpha}}_1{\vec{\phi}}+iw^2{\vec{\beta}}_1\partial{\vec{\phi}}}(w)>
\nonumber \\
+
{\vec{\alpha_1}}{\vec{\beta_1}}
<\partial^2{c}ce^{i{\vec{\alpha}}_1{\vec{\phi}}+i{\vec{\beta}}_1\partial{\vec{\phi}}}
(0)ce^{i{\vec{\alpha}}_1{\vec{\phi}}+iw^2{\vec{\beta}}_1\partial{\vec{\phi}}}(w)>
\nonumber \\
+{{\beta^2}\over{12}}
<\partial^3{c}ce^{i{\vec{\alpha}}_1{\vec{\phi}}+i{\vec{\beta}}_1\partial{\vec{\phi}}}
(0)ce^{i{\vec{\alpha}}_1{\vec{\phi}}+iw^2{\vec{\beta}}_1\partial{\vec{\phi}}}(w)>
\rbrack
\end{eqnarray}
First of all, as it is clear from (2.7) this correlator is only 
well-defined in case if the constraint 
\begin{equation}
{\vec{\alpha}}{\vec{\beta}}=0
\end{equation}
 is imposed.
Since for regular vertex operators $\alpha$ has a meaning of the momentum,
the orthogonality constraint (2.8) particularly implies that the $\beta$ parameter may be related to the
Fourier image of the extra coordinates in space-time in the context of double field theory and $T$-duality
(see the discussion section).

Furthermore, note that since the ghost correlator $<\partial^n{c}c(z_1)c(z_2)>=0$ for $n>2$, combined 
with the constraint
(2.8) the only surviving  terms in the correlator (2.7) are those proportional to $\sim{\partial{c}}c$ in 
$Q\Psi$.
In addition, in the on-shell  limit $\alpha_0^2\rightarrow{2}$ the correlators involving the terms 
$\sim\partial{c}c{\vec{\beta}}{{\partial}\over{\partial{\vec{\beta}}}}$ 
and $\sim\partial^2{c}c$ are of the order of $\sim{1\over{w}}$ and vanish.

Thus the only contributing correlator in the kinetic term gives

\begin{eqnarray}
<<Q\Psi(0)\star\Psi(0)>>
\nonumber \\
=
lim_{w\rightarrow\infty}
\int{d^D\alpha_1}{d^D\beta_1}\int{d^D\alpha_2}{d^D\beta_2}T({\vec{\alpha}}_1,{\vec{\beta}}_1)
T({\vec{\alpha}}_2,{\vec{\beta}}_2)w^{\alpha_2^2-1}e^{-{{{\vec{\alpha_2}}{\vec{\beta_2}}}\over{z}}}
\nonumber \\
\times\lbrace
({1\over2}\alpha_1^2-1+{\vec{\beta}}
{{\partial}\over{\partial{\vec{\beta}}}})<\partial{c}
ce^{i{\vec{\alpha}}_1{\vec{\phi}}+i{\vec{\beta}}_1\partial{\vec{\phi}}}
(0)ce^{i{\vec{\alpha}}_1{\vec{\phi}}
+iw^2{\vec{\beta}}_1\partial{\vec{\phi}}}(w)>\rbrace
\nonumber \\
=
\int{d^D\alpha}{d^D\beta}{1\over2}(\alpha^2-1)e^{\beta^2}T({\vec{\alpha}},{\vec{\beta}})
T(-{\vec{\alpha}},-{\vec{\beta}})
\end{eqnarray}
where we used the orthogonality constraint (2.8).
This concludes the computation of the rank 1  contribution to the kinetic term
in the SFT equations of motion .
Note that, in the regularity limit $\beta^2\rightarrow{0}$ (coinciding with the on-shell limit for the rank
1 irregular operator), one can expand the exponent so that the kinetic term in the Lagrangian
becomes
\begin{eqnarray}
\sim{\int}d\alpha{d}\beta{T}(-{\vec{\alpha}},-{\vec{\beta}})({1\over2}\alpha^2+\beta^2-1)T({\vec{\alpha}},
{\vec{\beta}})+...
\nonumber \\
\sim\int{d}xdyT(x,y)({1\over2}\Box_x+\Box_y+1)T(x,y)+...
\end{eqnarray}
where we skipped the higher derivative terms.
The next step is to calculate the cubic terms in the SFT equations.
We have:
\begin{eqnarray}
<<\Psi\star\Psi\star\Psi>>=\int\prod_{j=1}^3d\alpha_jd\beta_jT({\vec{\alpha}}_j,{\vec{\beta}}_j)
<g_j^3\circ{cqe^{i{\vec{\alpha_j}}{\vec{\phi}}+i{\vec{\beta_j}}{\vec{\partial{\phi}}}}}(0)>
\end{eqnarray}
Evaluating  the values of $g_j^3$ and their Schwarzian derivatives at $0$ and substituting the 
transformation laws
for $\Psi$ under $g_j^3$, as well as the on-shell constraints on ${\vec{\alpha}}$,
 it is straifghtforward to calculate:
\begin{eqnarray}
<<\Psi\star\Psi\star\Psi>>=\int\prod_{j=1}^3d\alpha_jd\beta_jT({\vec{\alpha}}_j,{\vec{\beta}}_j)
=
e^{{{5}\over{54}}(\beta_1^2+\beta_2^2+\beta_3^2)}(-{2\over3})^{{1\over2}\alpha_1^2-1}
(-{8\over3})^{{1\over2}\alpha_2^2+{1\over2}\alpha_3^2-2}
\nonumber \\
\times
<e^{{i{\vec{\alpha_j}}{\vec{\phi}}-{{2i}\over3}{\vec{\beta_j}}{\vec{\partial{\phi}}}}}(0)
e^{{i{\vec{\alpha_j}}{\vec{\phi}}-{{8i}\over3}{\vec{\beta_j}}{\vec{\partial{\phi}}}}}({\sqrt{3}})
e^{{i{\vec{\alpha_j}}{\vec{\phi}}-{{8i}\over3}{\vec{\beta_j}}{\vec{\partial{\phi}}}}}(-{\sqrt{3}})>
>
\nonumber \\
=
\int\prod_{j=1}^3d\alpha_jd\beta_jT({\vec{\alpha}}_j,{\vec{\beta}}_j)
\nonumber \\
exp\lbrace
{{{5}\over{54}}}(\beta_1^2+\beta_2^2+\beta_3^2)
+{{16}\over9}({\vec{\beta}}_1{\vec{\beta}}_2+{\vec{\beta}}_1{\vec{\beta}}_3+{\vec{\beta}}_2{\vec{\beta}}_3)
+{4\over{3{\sqrt{3}}}}({\vec{\alpha}}_2{\vec{\beta}}_3
\nonumber \\
-{\vec{\alpha}}_3{\vec{\beta}}_2)
+{2\over{3{\sqrt{3}}}}(4{\vec{\alpha}}_1({\vec{\beta}}_3-{\vec{\beta}}_2)+{\vec{\beta}}_1({\vec{\alpha}}_3
-{\vec{\alpha}}_2)
)\rbrace\delta(\sum_j\beta_j)\delta(\sum_j\alpha_j)
\nonumber \\
=
\int\prod_{j=1}^2d\alpha_jd\beta_jT({\vec{\alpha}}_1,{\vec{\beta}}_1)
T({\vec{\alpha}}_2,{\vec{\beta}}_2)T(-{\vec{\alpha}}_1-{\vec{\alpha}}_2,-{\vec{\beta}}_1-{\vec{\beta}}_2)
\nonumber \\
exp\lbrace
{-{{43}\over{27}}}({\vec{\beta}}_2{\vec{\beta}}_3+\beta_2^2+\beta_3^2)
+{2\over{{\sqrt{3}}}}({\vec{\alpha}}_2{\vec{\beta}}_3-{\vec{\alpha}}_3{\vec{\beta}}_2)\rbrace
\end{eqnarray}
Comparing the two-point and the three-point correlators, the irregular ansatz (1.9) solves the OSFT  
equation
of motion provided that the wavefunction $T_({{\vec{\alpha}},{\vec{\beta}}})$ satisfy the Euler-Lagrange 
equation following
from the cubic nonlocal effective action:
\begin{eqnarray}
S=-\int{d^Dx}{d^Dy}{\lbrace}T(x,y)e^{-\Box_y}(-{1\over2}{\Box_x}-1)T(x,y)
+{\tilde{\star}}\lbrace\tau^3(x,y)\rbrace
\end{eqnarray}
where 
\begin{eqnarray}
\tau(x,y)=e^{-{{43}\over{27}}\Box_y}T(x,y)
\end{eqnarray}
is a new (nonlocal) field variable, familiar from rolling tachyon cosmology
and the star product with the tilde is defined according to
\begin{eqnarray}
{\tilde{\star}}\lbrace{T_1(x,y)...T_N(x,y)}\rbrace
=lim_{y_1,...,y_N\rightarrow{\lbrace}
{y}}e^{\sum_{i,j=1;i<j}^N{{43}\over{27}}\partial_{y_i}\partial_{y_j}}
T(x,y_1)...T(x,y_N)\rbrace
\end{eqnarray}

This defines the analytic open string field theory solution in terms of rank one irregular 
vertex operators,
generating the higher-spin vertices on the leading Regge trajectory.
The generating wavefunction for higher spins is thus described, in the leading order, by the nonlocal 
action (2.13).
The actions of the type (2.13) are well known, as they describe extensions of
rolling tachyon dynamics \cite{senf, sens}, relevant to cosmological models with phantom fields 
.
The nonlocality coefficients appearing in the analytic solution (1.9), (2.13) must be related to 
cosmological parameters
of these models, such as dark energy state parameter and the vacuum expectation values
of the rolling tachyon in the equilibrium limit (with the
SFT solution interpolating between two vacua, describing the one dressed tachyon's value
$\tau\sim{e^{const\Box}}T$ evolving into the vacuum state satisfying the Sen's conjecture
constraints
\cite{senc, sencs}. 
The solution (1.9), (2.13) also defines the deformations of the BRST charge; solving the OSFT equations
with the deformed charge would then result in quartic and higher order corrections in
$\tau$. In the commutative level, nonlocal cosmological models of that type
have been considered in a number of works (e.g. see \cite{koshf, koshs, cosht, jouk}). 
In the case ${\vec{\beta}}=0$ (the regular case with the higher spins decoupled) the 
solution (1.9), (2.13) simplifies 
and  is described by the local cubic action, which is just the leading order low-energy effective
action for a tachyon in string perturbation theory. The solution (1.9), (2.13) is then the elementary
one, describing the perturbative background deformation of flat target space 
in the leading order of the tachyon's $\beta$-function.
With ${\beta}$-parameter switched on, the higher-spin dynamics enters the game and the effective
action becomes non-local, describing $non-perturbative$ background deformation in open string field theory.
The rank one solution, considered so far, can be understood as the one describing generating wavefunction
for higher-spin operators on the leading trajectory.
It is then straightforward to extend this computation to describe the SFT solutions involving the 
irregular 
blocks of higher ranks, generating the higher-spin vertices on arbitrary Regge trajectories.
The the effective action describing the generating higher-spin wavefunction essentially remains the same:
in the leading order, it is cubic in $\tau=e^{\sum_j=1^qa_j\Box_{\beta_j}}T({\vec{\alpha}},
{\vec{\beta}}_1,...{\vec{\beta}}_q)$
( $a_j$ are the constants defining the OSFT solution)
with the structure

\begin{eqnarray}
\sim\int{d\alpha}\prod_j{d{\beta}_j}
e^{\sum_j=1^qb_j\Box_{\beta_j}}T({1\over2}\alpha^2-1)T+{\tilde{\star}}(\tau^3)
\end{eqnarray}

where $\tau$ is again related to $T$ through nonlocal field redefinition.

All the family of the effective actions for collective
higher-spin wavefunctions is essentially nonlocal.
 Clearly, they must be related to the nonlocalities and the star products 
appearing
in higher-spin theories and Vasiliev's equations.
It is also remarkable that the generating wavefunction 
for higher spin fields thus emerges in the context of the 
rolling tachyon cosmology. Since the solutions of the type (1.9), (2.13) generally
describe the nonperturbative deformation of the flat background to collective
higher-spin vacuum, it is a profound question whether such a deformation, 
related to cosmological evolution of generalized rolling tachyon type objects,
is subject to constraints set up by the Sen's conjecture \cite{senc, sencs}.

\section{\bf Conclusion and Discussion}

In this work we have described simple analytic solution in open string
field theory,expressed in terms of vertex operators for irregular 
conformal blocks in the free limit of Toda theory, or bosonic string theory.
 The wavefunctions $T({\vec{\alpha}};{\lbrace}{\vec{\beta}}\rbrace)$ for 
these vertex operators are naturally related to those for the higher-spin vertex operators in 
open string theory 
according to
\begin{eqnarray}
H_{\mu_1...\mu_s}({\vec{\alpha}})
\sim(-1)^s\prod_{j=1}^s{{\delta^{n_j}}\over{\delta\beta^{\mu_j}_{n_j}}}
T({\vec{\alpha}};{\lbrace}{\vec{\beta}}\rbrace)\delta({\lbrace}{\vec{\beta}}\rbrace)
\end{eqnarray}
We found that the  irregular vertex operators analytically solve the equations of open string field theory,
provided that the generating wavefunctions satisfy the constraints  following from the cubic effective 
action
for generalized rolling tachyons. This implies that the ``engineering'' of nonperturbative 
higher-spin vacuum
(the nonperturbative deformation of the background from flat to the one described by the minimum of 
collective
higher-spin action, computed to all orders) can be mimicked by the evolution of a rolling tachyon
interpolating between inequivalent vacua in cosmological models for the dark energy, with the nonlocality
introduced in order to approach the ``Big Rip'' problem that occurs when the equation of state parameter
 $\sigma$
 in the equation of state $p=\sigma\rho$ (with $p$ and $\rho$ being the pressure and 
the energy densities) is less then $-1$.
To understand the interplays between higher spin dynamics and cosmological models, it shall be  
important to 
establish the explicit form of the OSFT solution for higher irregular ranks, in order to include the 
higher spins on subleading trajectories,
and to analyze the equations of motions for the  generating wavefunctions $T({\vec{\alpha}},
\lbrace{\vec{\beta}}\rbrace)$.
We hope to elaborate on this in the future work, currently in progress.

As the $\beta$-parameters entering the generating wavefunctions are related to the higher-spin couplings,
 it is 
also worth commenting their physical meaning in the context of global space-time symmetries.
The $\alpha$-parameter is clearly related to the momentum of the regular  part of the vertex operators.
Since $L_0U_N={1\over2}(\alpha^2+...)U_N$, the Virasoro generator 
$L_0=-{1\over2}\oint{dz}zP_m^2$ where $P_m=\partial{X}_m$ is the current for space-time translations,
is obviously a quadratic Casimir. As the number of the Toda field (target space) components 
increases, 
so does the number of Casimir operators of the space-time symmetry algebra, as well as the highest 
possible ranks of
 $W_n$ currents with the irregular vertices being their eigenvectors.
Roughly, the highest $W_n$ rank grows with the number  Casimirs, but this correspondence is 
in not an exact match for $n>4$  and rather subtle, related to an unsolved problem in number theory
regarding the calculation of the number of ordered partitions of a given length - in general, 
the rank of $W_n$ grows
faster than the rank of space-time symmetry algebra. Does this imply the presence of hidden dimensions
 in the  theory?
In case of the rank one, the single ${\vec{\beta}}$-parameter also seems to have an interesting 
relation to the
$T$-duality transformations in the double field theory context, suggested by orthogonality relation (2.8).
Indeed, suppose $\phi$ is a target space coordinate, compactified on a circle. Then $\partial\phi$ is 
an operator
for an infinitezimal the radius change, while $e^{\beta\partial\phi}$ would define the finite deformations
of the compactified dimension. This would in turn define the $T$-duality transformation with the 
compactification radius
$R\sim\beta^{-1}+{\sqrt{4+\beta^{-2}}}$. The explicit construction of vertex operators in double 
string theories
could then be realized in terms of operator algebras involving irregular operators acting on 
regular states.
We hope to address this issue, as well as those outlined above, in the future works.

\begin{center}
{\bf Acknowledgements}
\end{center}

The author acknowledges the support of this work  by the National Natural 
Science Foundation of China under grant 11575119.


\end{document}